# Snapshot multi-spectral imaging through defocusing and a Fourier imager network

## Authors

Xilin Yang[1,2,3]*, Michael John Fanous[1]*, Hanlong Chen[1,2,3]*, Ryan Lee[1], Paloma Casteleiro Costa[1,2,3], Yuhang Li[1,2,3], Luzhe Huang[1,2,3], Yijie Zhang[1,2,3], Aydogan Ozcan[1,2,3,§]

[1]Electrical and Computer Engineering Department, University of California, Los Angeles, CA, 90095, USA.

[2]Bioengineering Department, University of California, Los Angeles, CA, 90095, USA.

[3]California NanoSystems Institute (CNSI), University of California, Los Angeles, CA, 90095, USA.

*Equal contribution

§Corresponding author: ozcan@ucla.edu

## Abstract

Multi-spectral imaging, which simultaneously captures the spatial and spectral information of a scene, is widely used across diverse fields, including remote sensing, biomedical imaging, and agricultural monitoring. Here, we introduce a snapshot multi-spectral imaging approach employing a standard monochrome image sensor with no additional spectral filters or customized components. Our system leverages the inherent chromatic aberration of wavelength-dependent defocusing as a natural source of physical encoding of multi-spectral information; this encoded image information is rapidly decoded via a deep learning-based multi-spectral Fourier Imager Network (mFIN). We experimentally tested our method with six illumination bands and demonstrated an overall accuracy of 92.98% for predicting the illumination channels at the input and achieved a robust multi-spectral image reconstruction on various test objects. This deep learning-powered framework achieves high-quality multi-spectral image reconstruction using snapshot image acquisition with a monochrome image sensor and could be useful for applications in biomedicine, industrial quality control, and agriculture, among others.

## Introduction

Multi-spectral imaging, with the capacity of concurrently capturing the spatial and spectral information of objects, has become an indispensable tool in various fields, including remote sensing[1–3], biomedical imaging[4–7], and agricultural and food monitoring[8–12], among others[13,14].

However, unlike standard color image sensors that utilize a Bayer filter array to capture the red, green, and blue (RGB) channels, multi-spectral imaging typically requires bulkier optical set-ups to separately capture spatial data within each wavelength channel while minimizing spectral crosstalk among the channels of interest. One of the approaches for multi-spectral imaging involves scanning systems with single-pixel sensors or one-dimensional sensor arrays coupled with wavelength-selective components like color filters[15,16] or prisms[17]. However, these scanning systems acquire spatial and spectral data sequentially, which compromises frame rate, increases exposure time, and degrades signal-to-noise ratio, thereby limiting their applicability in real-time imaging scenarios. Non-scanning systems, commonly referred to as snapshot imagers[18], aim to capture spatial and spectral information in a single exposure. Examples of snapshot imagers include color-coded apertures[19], Fabry-Pérot filters[20,21], spectral filter arrays[22,23], metasurfaces[24,25] and diffractive optical elements[26,27,28]. Despite their ability to perform real-time imaging, most snapshot multi-spectral imaging systems require additional components that are relatively complex and less accessible compared to commercially available image sensors used in everyday applications.

Here, we present a single-shot multi-spectral imaging system that utilizes a simple monochrome image sensor without any spectral filters or customized components. For the multi-spectral image encoding process, we leverage chromatic dispersion and image defocusing to serve as our physical encoder. For the multi-spectral image decoding, we employ a trained neural network termed multi-spectral Fourier Imager Network (mFIN), which is optimized to rapidly reconstruct the multi-spectral image information embedded in the defocused image captured using a monochrome image sensor. We experimentally validated our approach by acquiring 964 multi-spectral images, where each object is acquired by illuminating a pattern with a distinct combination of light-emitting diodes (LEDs). The entire dataset was divided into 888 multi-spectral objects used to train the mFIN model and the remaining 76 multi-spectral objects, unseen during training, were employed to blindly evaluate the model's ability to accurately reconstruct images and their corresponding color channels. Our results demonstrate that this multi-spectral imaging system achieved an overall accuracy of 92.98% in predicting the correct illumination channels of the test objects using a monochrome image sensor array. Additionally, the reconstructed multi-spectral images of test objects exhibit a decent structural quality, as evidenced by various quantitative image metrics including the Structural Similarity Index Measure (SSIM) averaging $0.63 \pm 0.09$, Peak Signal-to-Noise Ratio (PSNR) averaging $12.00 \pm 1.82$ dB, Root Mean Square Error (RMSE) averaging $0.26 \pm 0.06$, and Normalized Mean Squared Error (NMSE) averaging $0.068 \pm 0.026$. These experimental results confirm the effectiveness of our snapshot computational multi-spectral imaging method using a monochrome image sensor. Our study presents an efficient multi-spectral imaging solution that integrates natural image defocusing as a physical encoding scheme with a deep learning-based spectral decoding framework. This integration not only simplifies multi-spectral imaging hardware but also maintains high-quality image reconstructions through snapshot image acquisition, potentially enabling broader applications and enhancing the accessibility of multi-spectral imaging technologies.

## Results

**Multi-spectral imaging with image defocus-based physical encoding**

Figure 1 illustrates our system set-up, depicting both the schematic layout (Fig. 1a) and the physical components of the system (Fig. 1b). To achieve multi-spectral imaging, we implemented a custom-designed ring-shaped LED array comprising six different types of LEDs that span the visible spectrum. The multi-spectral illumination is directed through a collimator, diffuser, and pinhole, ultimately reaching a digital micromirror device (DMD). The DMD projects the spatial information of the image, referred to as the image pattern, thereby forming the final multi-spectral image. A 3× demagnifying lens pair then focuses this multi-spectral image onto a specific region of a monochrome complementary metal-oxide-semiconductor (CMOS) image sensor. Due to the inherent chromatic aberration of the optical elements and dispersion, different illumination wavelengths result in varying focal planes. Consequently, our system captures each spectral channel with a distinct level of image defocus and blur, resulting in physical defocus-based multi-spectral image encoding. This defocused and blurred pattern that is captured by a monochrome image sensor is subsequently decoded by a deep learning-based neural network model, enabling the reconstruction of the multi-spectral image of an unknown pattern. This multi-spectral image reconstruction is performed using a multi-spectral Fourier Imager Network termed mFIN, which features residual connections with channel repetition from the monochrome image input to the multi-spectral image output, ensuring the alignment of the channel dimensions (see the Methods section for details).

A total of 964 objects from 120 unique image patterns of test objects (images of human lung tissue samples; see the Methods section) were imaged, each captured under a distinct illumination condition using various combinations of six LEDs. Of these, 888 multi-spectral objects from 114 patterns were utilized in the training process of the mFIN model. The training involved optimizing the mFIN network by minimizing the discrepancies between the ground truth image patterns (projected by the DMD) with channels illuminated by their corresponding LEDs that were turned on and the multi-spectral images reconstructed by the mFIN network. Further details regarding the dataset and training methodology are provided in the Methods section.

Following the training process, we experimentally evaluated our model using blind testing data, consisting of 76 multi-spectral objects of 6 test patterns that were never encountered during the training phase. We first evaluated our system's ability to accurately reconstruct images of test objects and their corresponding color channels under a single-LED illumination, with the results presented in Figure 2. The left column displays the snapshot grayscale images (resulting from the monochrome image sensor), which exhibit varying degrees of defocus, aberrations, and intensity fluctuations across different illumination wavelengths. For each test object, mFIN accurately predicted both the corresponding illumination wavelength and the underlying image pattern, where each illumination channel was denoted by the center wavelength of its corresponding LED; see Figure 2.

In the subsequent phase of the blind testing, we increased the complexity of the multi-spectral image reconstruction task by testing new image patterns under simultaneous illumination by multiple LEDs, covering various combinations of wavelengths. Figure 3 showcases the test image patterns illuminated by different LED configurations. The input monochrome images captured by the CMOS imager (top row) exhibit significant spatial variations and aberrations due to the superposition of multiple illumination wavelengths. mFIN reconstructed output images (left column) are juxtaposed with the target images (right column), revealing a high degree of concordance between the two. These experimental results further confirm that mFIN can blindly generate accurate multi-spectral image reconstructions from snapshot monochrome images, demonstrating its robustness.

To quantitatively evaluate the performance of our model, we first assessed its spectral channel accuracy using a classification-based approach, followed by an analysis of the structural image metrics for the spatial quality of the reconstructed multi-spectral image features. For the spectral channel accuracy classification, each illumination LED channel was treated as a binary state: the LED was either on during the test object illumination, contributing to the snapshot monochrome image, or off, leaving the spectral channel of this object blank. Consequently, spectral prediction errors were categorized into two types: false positives (termed "leakages"), where the model incorrectly predicted an image in a blank/off spectral channel, and false negatives (termed "losses"), where the model failed to generate a spectral image or produced one with low intensity in a spectral channel that should contain an image. The accuracy of the spectral predictions was determined by applying thresholds on an energy difference term defined as $I_{dif} = \frac{I_{pred} - I_{target}}{I_{target}}$ where $I_{pred}$ represents the predicted image intensity of a given output channel, calculated as the sum of all the pixel values in the predicted image and $I_{target}$ denotes the sum of the pixel intensity values of the ground truth image pattern (see Figure 4a and the Methods section for details). For statistical analysis, each sample pattern under a given illumination configuration was treated as six independent entries—one per channel—resulting in a total of 456 data points derived from 76 test objects never seen during training. Our multi-spectral image reconstruction model reached an overall spectral accuracy of 92.98% across all data points in the test dataset. The confusion matrix, presented in Figure 4a, provides a detailed breakdown of the spectral prediction results (also see the Methods section). While the false negative rate (losses) was slightly higher than the false positive rate (leakages), both error types remained within acceptable levels, as shown in Figure 4a.

Next, we analyzed the model's performance relative to the number of concurrent illuminations by calculating sensitivity, specificity, and F1 score values (see the Methods section) across varying illumination configurations. As anticipated, the prediction performance slightly decreased with an increasing number of concurrent illuminations. For the five-LED configuration, we observe the predominance of active LEDs, which resulted in very few channels/data points with negative ground truth. Consequently, the scarcity of true negative samples led to a relative reduction in specificity (see Figure 4b).

While classifying the spectral predictions as correct or incorrect (Figures 4a-b) provides a binary assessment of our performance, it does not fully capture the nuances of the reconstructed image quality. To address this, Figure 4c presents the energy difference levels for all test data points using boxplots, categorized by the number of concurrent illuminations. For enhanced visibility, a zoomed-in version of the y-axis is also shown on the right. The results reveal that as the number of concurrent LED illuminations increases, the mean energy difference gradually shifts away from zero, accompanied by a steadily increasing standard deviation. This trend indicates that the performance of the multi-spectral image reconstructions decreases as the number of simultaneous LED illuminations grows, likely due to the increasing complexity of the superimposed illumination patterns captured by a monochrome image sensor. Despite this, most data points still fall within a narrow range of ±0.2 in the energy difference. This consistent performance within a tolerable range of error demonstrates the reliability and effectiveness of our approach in reconstructing multi-spectral images under more challenging conditions.

Next, we analyzed how the wavelength of the illumination affects the multi-spectral image reconstruction performance. Similar to the earlier analyses, each test image from a single spectral channel was treated as an independent data point, resulting in six data points per test pattern—one for each spectral channel. We then grouped these data points by their corresponding illumination wavelengths and plotted the energy difference levels in Figure 4d. This analysis further reveals the variability in image reconstruction performance across different wavelengths. For instance, the 532 nm illumination wavelength tends to exhibit a higher rate of false positives, where the model incorrectly predicts an image in a blank channel, resulting in "leakages". Conversely, the 605 nm wavelength shows a higher rate of false negatives, where the model fails to predict an image that should be present, resulting in "losses". These discrepancies can be attributed to variations in how each wavelength interacts with the optical system, particularly through chromatic aberrations and focal plane shifts during the physical defocus-based multi-spectral image encoding process. We also evaluated the quality of the reconstructed images using various structural image metrics, including RMSE, SSIM, PSNR, and NMSE (refer to the Methods section for details). When assessing these metrics, we focused exclusively on true positive predictions, as leakages or losses cannot be used for quantitative evaluation of image quality. In its blind testing, our model achieved an average SSIM of $0.63 \pm 0.09$ and a PSNR of $12.00 \pm 1.82$ dB. Additionally, the RMSE averaged $0.26 \pm 0.06$, and the NMSE metric averaged $0.068 \pm 0.026$ for the test objects. These results, characterized by low standard deviations, demonstrate a repeatable performance in our model's multi-spectral image reconstructions. In Figure 5a–d, we further report the distributions of these four quality metrics across different wavelengths. The results are consistent across all four image quality metrics: the reconstructions at 460 nm and 623 nm perform better in general, in line with the earlier spectral accuracy measurements. In contrast, the illumination wavelengths 397 nm and 590 nm exhibit slightly lower performance, although still acceptable. Finally, we evaluated the reconstructed multi-spectral image quality and the resulting confusion matrices in relation to the number of concurrent LED illuminations, as depicted in Figure 6. The performance remained relatively stable across varying levels of multiplexed illumination. Despite minor discrepancies observed, the presented snapshot multi-spectral imaging system performs well across all the tested wavelengths, as indicated by the quantitative structural image quality metrics, confirming the robustness of the reconstruction process across the visible spectrum.

**Discussion**

The snapshot multi-spectral imaging system presented in this study demonstrates promising capabilities in reconstructing image patterns with high accuracy and quality across multiple illumination wavelengths using an image-defocusing and aberration-based encoding mechanism combined with a deep learning-based decoding framework. However, the current experimental implementation is subject to hardware constraints due to the DMD that is used for training and testing our system. As the inverse problem becomes increasingly more complex, i.e., when addressing multi-spectral images with finer structural details and higher bit-depth patterns for each spectral channel, image-defocusing-based encoding of spectral information may not sufficiently capture all the information needed for high-fidelity multi-spectral image reconstructions. To provide better multi-spectral image encoding strategies, there is growing research interest in utilizing diffractive optical elements to engineer wavelength- and depth-dependent point-spread functions (PSFs), potentially enhancing how information is projected onto image sensors. When combined with deep neural network models, these engineered PSFs have the potential to significantly improve the quality of multi-spectral image reconstructions. For example, recent advancements have demonstrated that diffractive deep neural networks[29] can effectively design spatially and spectrally varying PSFs for incoherent light sources[30,31] with multiple diffractive layers interconnected through free-space light propagation. By incorporating such cascaded structured diffractive layers, one can replace the defocusing-based multi-spectral image encoding strategy with a fully programmable diffractive approach, providing greater degrees of freedom compared to traditional single-layer diffractive designs[30,32]. Therefore, the joint optimization of diffractive networks alongside back-end multi-spectral image reconstruction network models could yield a powerful solution, pushing the boundaries of what is achievable in multi-spectral and hyperspectral imaging systems. Such integrated optimization strategies of front-end optical encoder hardware and back-end digital decoder algorithms would allow for the simultaneous refinement of both the physical imaging set-up and the computational back-end system, ensuring that the entire imaging pipeline is finely tuned for multi-spectral image reconstruction fidelity and performance.

## Methods

**Network architecture, training and performance evaluation metrics**

mFIN architecture is composed of multiple Dynamic Spatial Fourier Transform (dSPAF) modules with dense connections[33], as shown in Fig. 7a. The dense connections, also referred to as dense links, aggregate the output of all previous dSPAF modules to the next one (Fig. 7b) so that the feature maps are progressively accumulated, allowing the network to retain information flow while effectively capturing intricate spatial representations. The dSPAF module is shown in Fig. 7c, utilizing a shallow U-Net[34] to dynamically generate spatial frequency filters, enabling adaptive processing and introducing extra degrees of freedom. Input images are first converted to the spatial frequency domain using a two-dimensional Fast Fourier Transform (FFT), which are then modulated by the dynamically generated weights from the shallow U-Net. Subsequently, an inverse Fast Fourier Transform (iFFT) is performed to reconstruct the image in the spatial domain. The resulting spatial tensors undergo a Parametric Rectified Linear Unit (PReLU)[35] activation to introduce non-linearity. A residual connection was incorporated to provide better convergence by establishing a direct pathway from input to output; channel repetition was also employed for the monochrome input to match the desired number of output channels.

Our training loss function is a weighted sum of the mean absolute error (MAE) term and the Fourier domain MAE (FDMAE) loss:

$$Loss_{emFIN} = \omega_{MAE} L_{MAE} + \omega_{FDMAE} L_{FDMAE}$$

where $\omega_{MAE}$ and $\omega_{FDMAE}$ are weights for each domain and are empirically set to be 1 and 0.015, respectively. Each loss term is defined as:

$$L_{MAE} = \frac{\sum_{i=1}^{n} \text{abs}(y_i - \hat{y}_i)}{n}$$

$$L_{FDMAE} = \frac{\sum_{i=1}^{n} \text{abs}(\mathcal{F}(y)_i - \mathcal{F}(\hat{y})_i)}{n}$$

where $\mathcal{F}$ defines the 2D discrete Fourier transform, $y$ represents the ground truth, $\hat{y}$ represents the output of the network, and $n$ is the total number of pixels.

We defined the energy difference for each channel as $I_{dif} = \frac{I_{pred} - I_{target}}{I_{target}}$ where $I_{pred} = \sum_i \hat{y}_i$ and $I_{target} = \sum_i Y_i$, where $i$ is the index of all pixels within the image. Note that we calculated $I_{target}$ using the binary image pattern ($Y$) irrespective of whether the channel was illuminated (positive ground truth) or blank (negative ground truth). This approach ensures that $I_{dif}$ can always be normalized without division by zero, even if the channel's ground truth appears blank (i.e., the corresponding LED is not turned on). We applied two thresholds on the energy difference values to determine whether a sample is considered correctly predicted: 0.2 for false negatives and 0.1 for false positives. More specifically, for a sample where the ground truth is positive, any output that has $I_{dif} \leq -0.2$ is considered a false negative or loss. Similarly, for a sample where the ground truth is blank, any output with $I_{dif} \geq 0.1$ is considered a false positive or leakage. These thresholds have been chosen empirically to balance the trade-off between the prediction accuracy and the overall image quality.

For performance evaluations, we used sensitivity, specificity and F1 score for overall channel accuracy metrics:

$$\text{sensitivity} = \frac{TP}{FN + TP}$$

$$\text{specificity} = \frac{TN}{FP + TN}$$

$$\text{F1 score} = 2 \times \frac{\bigl(TP/(TP+FP)\bigr) \times \bigl(TP/(TP+FN)\bigr)}{\bigl(TP/(TP+FP)\bigr) + \bigl(TP/(TP+FN)\bigr)}$$

where $TP, TN, FP, TN$ represent the counts of true positives, true negatives, false positives, and false negatives, respectively.

For reconstructed image quality evaluations, we used four structural metrics: RMSE, SSIM, PSNR and NMSE. Unless otherwise stated, we calculated all metrics on the raw predicted images without scaling, normalization, or binarization operations. RMSE is defined as:

$$\text{RMSE}(y, \hat{y}) = \sqrt{\frac{\sum_{i=1}^{n}(y_i - \hat{y}_i)^2}{n}}$$

where $y_i$ represents the ground truth image pixel, and $\hat{y}_i$ represents the output image pixel from the model, $i$ is the index for all pixels within the image. NMSE metric focuses on the image contrast and uses a normalization factor $\sigma$ to rule out the impact of absolute image intensity level mismatches, i.e.:

$$\text{NMSE}(y, \hat{y}) = \text{RMSE}(\sigma y, \hat{y})$$

where $\sigma = \frac{\sum_i \hat{y}_i}{\sum_i y_i}$.

SSIM evaluates the structural similarity between two images, defined by:

$$\text{SSIM}(y, \hat{y}) = \frac{(2\mu_y \mu_{\hat{y}} + c_1)(2\sigma_{y\hat{y}} + c_2)}{(\mu_y^2 + \mu_{\hat{y}}^2 + c_1)(\sigma_y^2 + \sigma_{\hat{y}}^2 + c_2)}$$

where $\mu_y$ and $\mu_{\hat{y}}$ represents the mean pixel values of the ground truth and the output images while $\sigma_y^2, \sigma_{\hat{y}}^2$ are the variances and $\sigma_{y\hat{y}}$ is the covariance of the ground truth and the output images. $c_1$ and $c_2$ are constants, which were calculated based on the range of the images to be ~6.5 and ~58.5, respectively.

PSNR is defined as:

$$\text{PSNR} = 20\log_{10}(\frac{1}{\text{RMSE}(y,\hat{y})})$$

The training and testing are done on a standard workstation with GeForce RTX 3090 Ti graphics processing units (GPU) in workstations with 256GB of random-access memory (RAM) and Intel Core i9 central processing unit (CPU). mFIN was implemented using Python version 3.12.0 and PyTorch[36] version 2.1.0 with CUDA toolkit version 12.1.

**Experimental set-up, data acquisition and pre-processing**

A custom-designed multi-spectral LED array was built using six distinct LEDs: red (620-625 nm), green (520-525 nm), purple (395-400 nm), yellow (588-592 nm), orange (600-610 nm), and blue (455-465 nm). The LEDs were arranged in a circular configuration on a 1" x1" prototype board, as shown in Figure 1. The array was modulated using an Arduino UNO R3 microcontroller board, programmed with the Arduino IDE 1.8.19 software[37] to manage both the cycling and brightness of the LEDs. The cycling pattern involved several phases to maximize the capture of spectral information. First, each LED was illuminated individually for 5 seconds. Following this, combinations of two LEDs were activated, cycling through various pairs. After the two-LED combinations, the system progressed to groups of three LEDs, illuminating them simultaneously for further spectral complexity. Further combinations of four and five LEDs were also included, providing a rich dataset for encoding by mixing multiple wavelengths. Each phase of the cycle (ranging from 1 to 5 LEDs for each combination) was executed for a designated duration, with brightness levels adjusted to account for varying LED photon energy and ensure roughly uniform brightness throughout our measurements. The Arduino controlled both digital and pulse width modulation pins, allowing for precise control of the intensity of each LED.

The optical system was designed to capture encoded multi-spectral information using a DMD and a custom camera set-up. A Texas Instruments DLP9500 DMD was used, featuring a 10.8 µm pitch size, and was driven by the EasyProj (ViALUX) software to project binary images of lung tissue samples. To counteract the DMD's 45-degree tilt, the image was displayed in a diamond shape against a black background. The multi-spectral illumination passed through a collimator and diffuser before striking the DMD and was subsequently focused by two lenses: one with a focal length of 200 mm and the other 75 mm, positioned 110 mm apart. The imaging was captured by a ScopeTek DCM500BW camera housed in a custom 3D-printed casing made using a Stratasys Objet30 Prime™ printer. The camera, equipped with a 5.0-megapixel monochrome image sensor (2592x1944 resolution, 2.2 µm pixel size), recorded out-of-focus images, which were processed and stored via the ScopePhoto 3.1 software.

Formalin-fixed, paraffin-embedded (FFPE) tissue slides stained with hematoxylin and eosin (H&E) from anonymized lung specimens were used for testing. Each slide originated from previously collected anonymized materials that did not contain identifiable patient information. The slides were initially imaged as holographic data using a custom-built lens-free in-line setup[38,39]. The reconstructed holographic images of tissue specimens were then binarized using Otsu's thresholding method[40], which adaptively selects the optimal threshold to separate foreground and background regions. The resulting images were rotated into a diamond shape and positioned against a black background to create an upright frame of suitable size on the CMOS image sensor. The optical configuration resulted in the captured images being approximately three times smaller than the projected images, due to the lens system and component distances. The image defocus varied between different illumination wavelengths due to chromatic aberration. Blue light focuses closer to the lens, while red light focuses farther away, leading to differences in the amount of defocus for different colors.

For the training and testing datasets, we collected two sets of image patterns. The first set consisted of 100 image patterns, each with 6 exposures corresponding to one unique LED illumination at a time. The second set consisted of another 20 image patterns. In addition to the 6 illumination settings in the first set, each projected image pattern in the second set is illuminated with another 14 illumination settings that have more than one LED on at a time. In total, we captured 964 multi-spectral images using these two image pattern sets. For blind testing purposes, we isolated 76 multi-spectral objects from 6 patterns – 3 from the first set and 3 from the second – as our testing dataset, used to test the generalization capability of our model. The captured raw monochrome images, encompassing the full field-of-view on the image sensor, underwent the following pre-processing steps: we first cropped the center diamond regions and rotated the images 45 degrees to roughly align the captured data to be in the same orientation with the target images projected to the DMD. A 5 × 5 pixel binning was applied to the input images, which were then resized to 512 × 512 pixels to ensure compatibility with the network model architecture. This resolution was fixed for both the training and evaluation phases to maintain consistency. All resizing operations used bilinear kernels.

**Figures and captions**

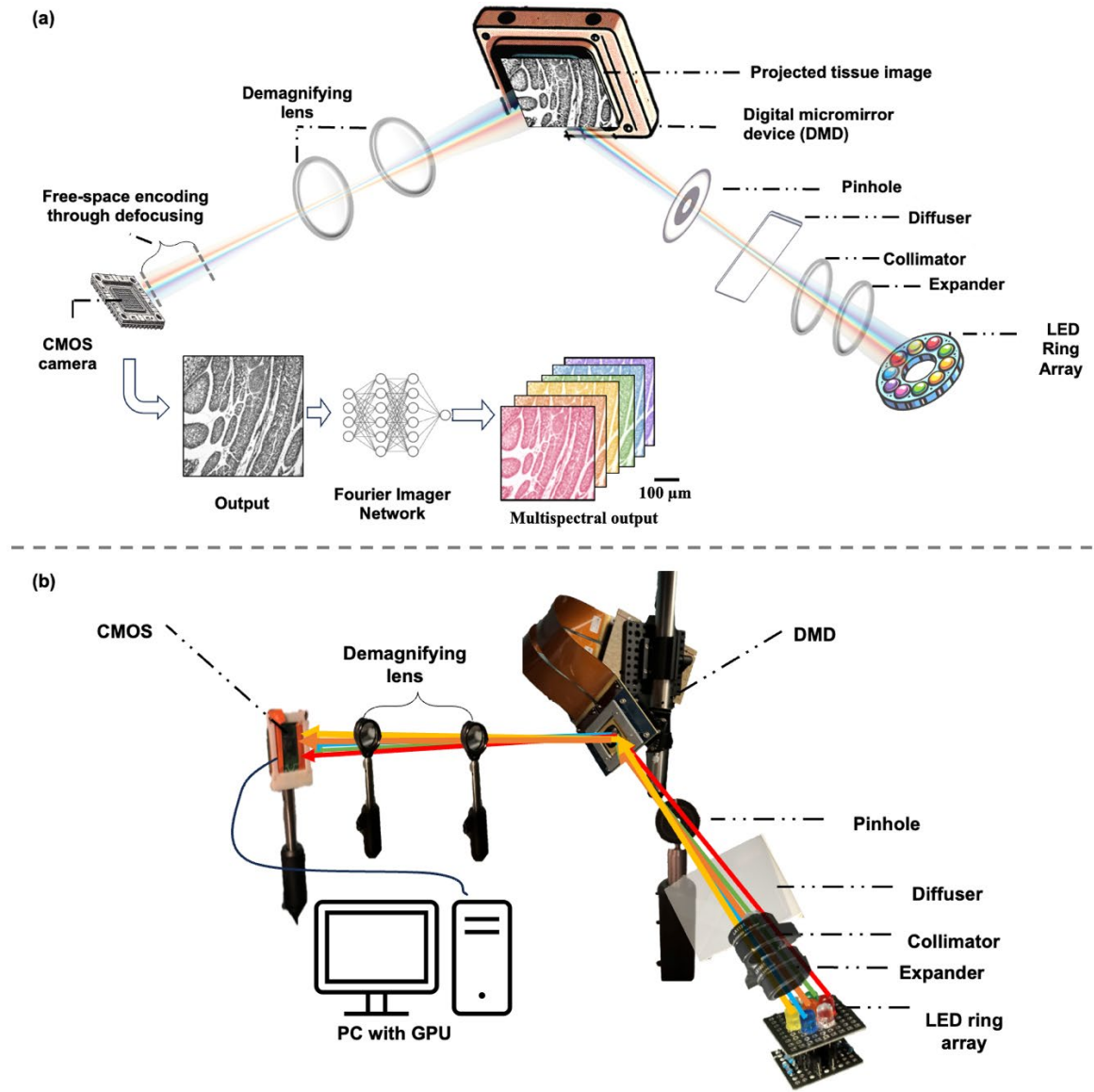

**Figure 1**. (a) Schematic layout: Overview of the snapshot multi-spectral imaging system, illustrating the arrangement of the circular LED array, Digital Micromirror Device (DMD), optical components (collimator, diffuser, lenses), and the monochrome imaging camera. (b) Experimental set-up.

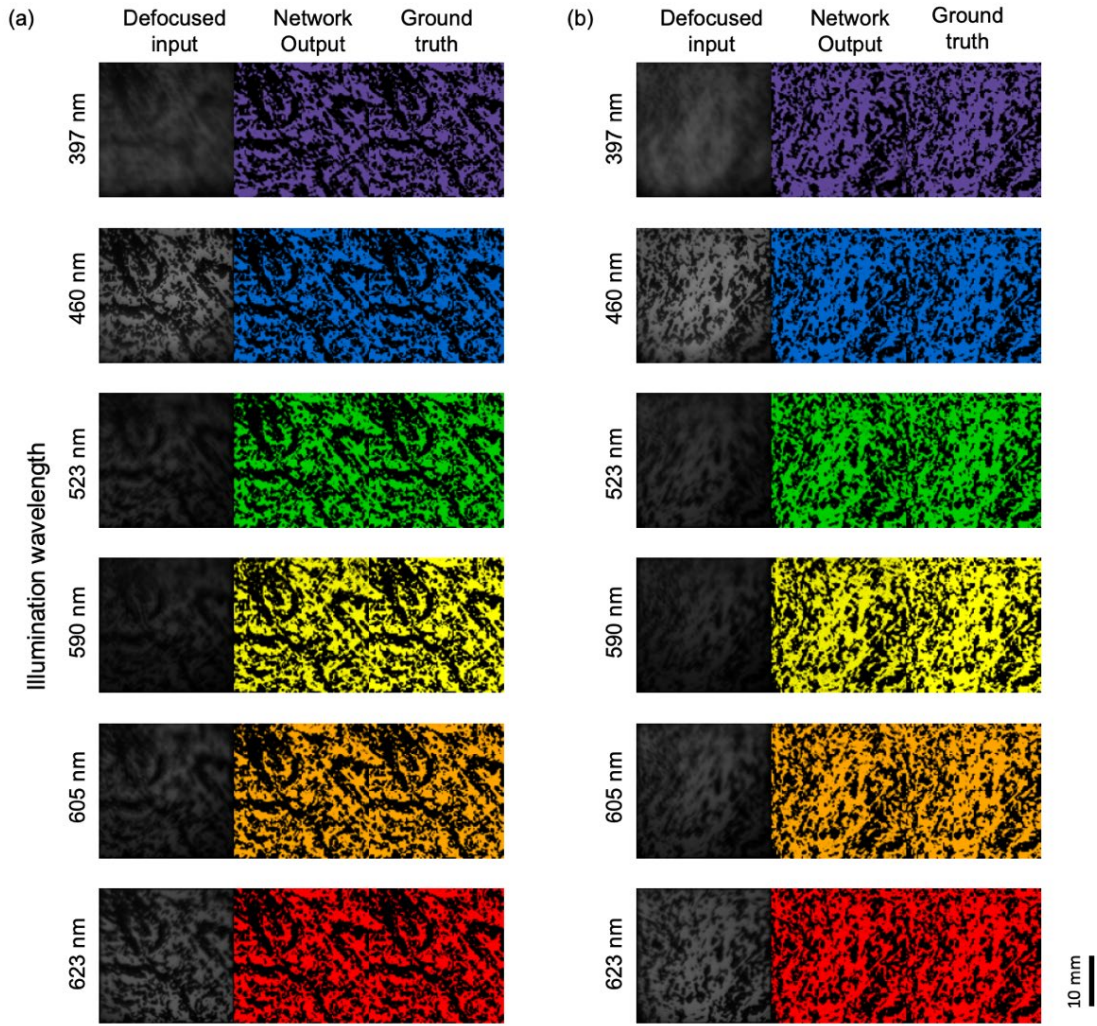

**Figure 2**. (a-b) Visualizations of image reconstruction results of two test image patterns illuminated under different wavelengths. Each column refers to an image with the monochrome defocused sensor input, network output and multi-spectral target (ground truth). Pseudo colors represent the corresponding illumination LED color (center wavelength).

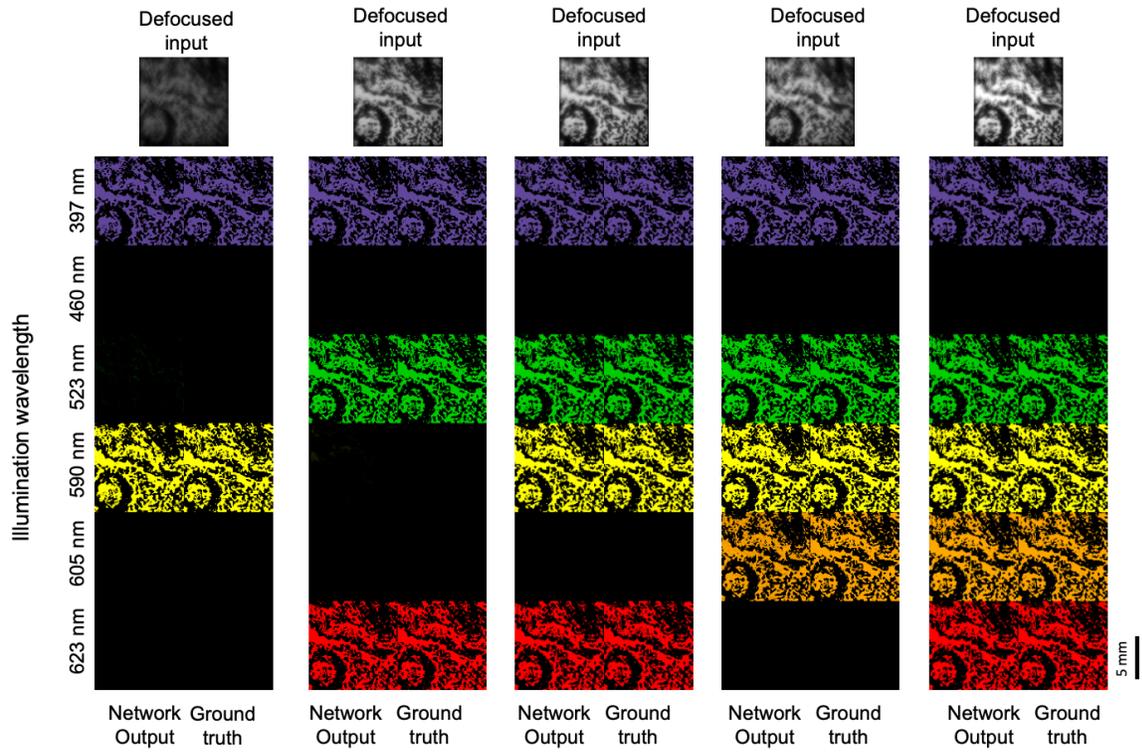

**Figure 3**. Visualization of image reconstruction results under various illumination configurations. Each column displays the predicted image alongside the corresponding target (ground truth) image for different combinations of activated LEDs that were on.

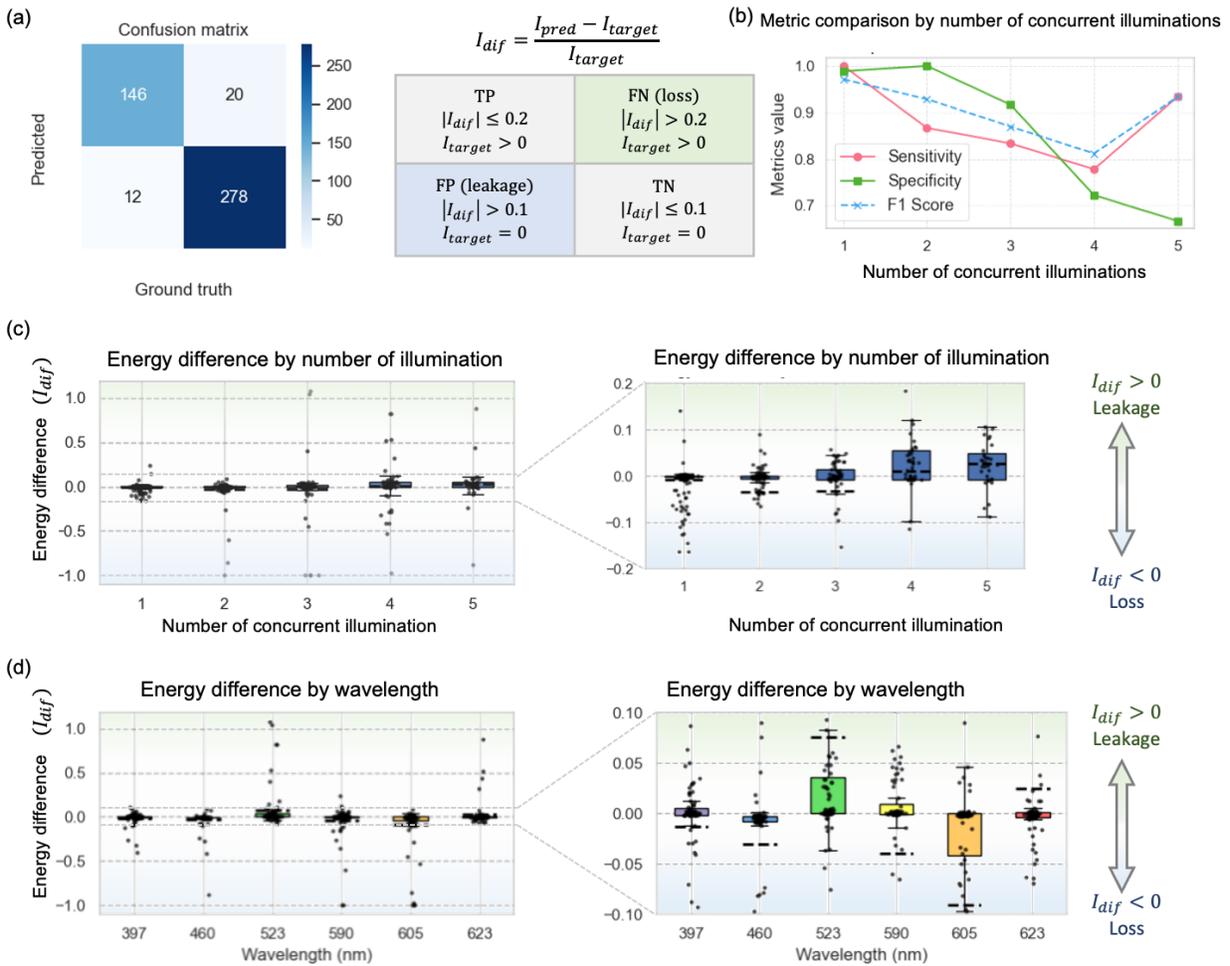

**Figure 4**. Image reconstruction performance analysis. (a) Confusion matrix of spectral channel accuracy. (b) Classification metrics against the number of concurrent LED illuminations: Plots of sensitivity, specificity, and F1 score against the number of concurrent LED illuminations. (c) Energy difference values as a function of the number of concurrent LED illuminations; boxplots depict the distribution of energy differences for various numbers of concurrent illuminations, indicating the shift in the prediction accuracy and variance. (d) Energy difference values as a function of the illumination wavelength. We used light green to signify "losses" and light blue for "leakages".

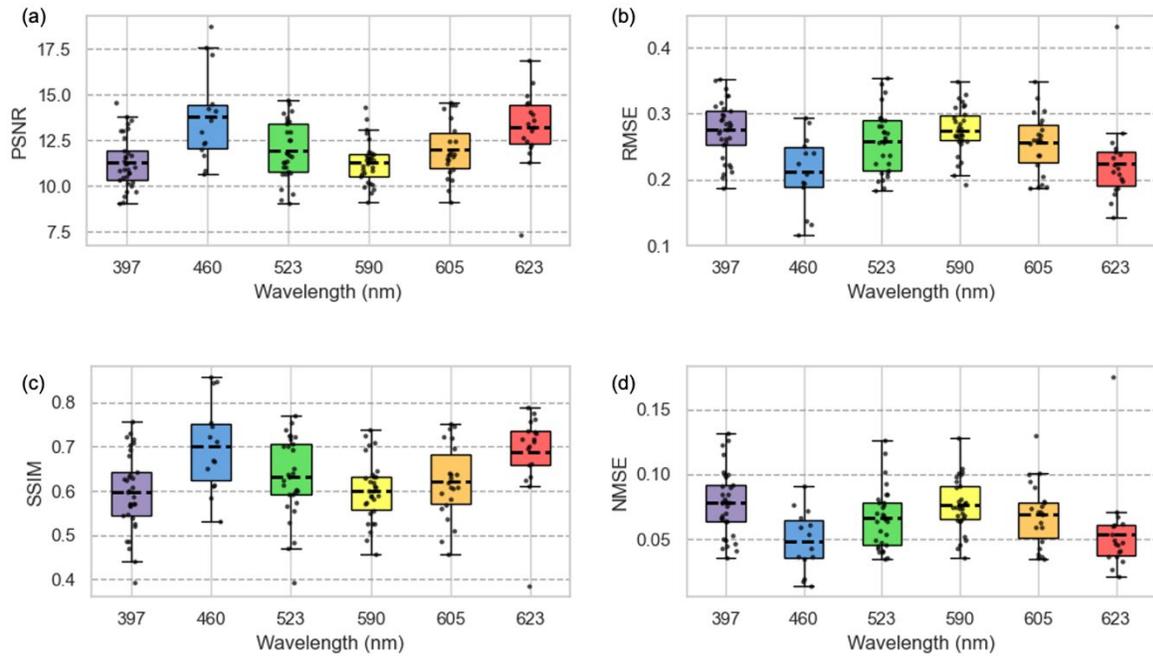

**Figure 5**. Distributions of (a) PSNR, (b) RMSE, (c) SSIM and (d) NMSE metrics for the reconstructed images at different illumination wavelengths.

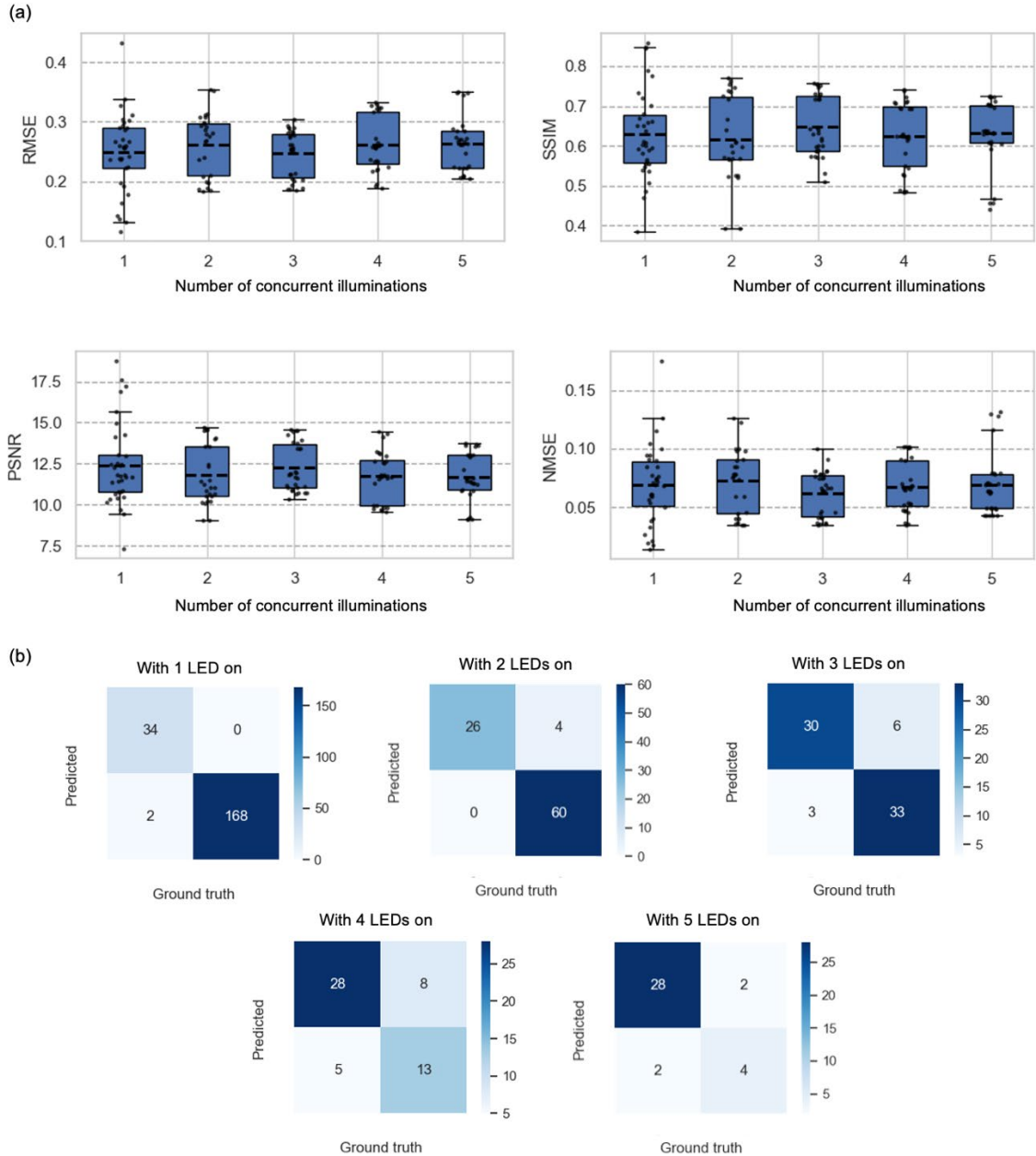

**Figure 6**. (a) Distributions of structural quality metrics for the reconstructed multi-spectral images with respect to the number of concurrent LED illuminations. (b) Confusion matrices with different numbers of concurrent LED illuminations.

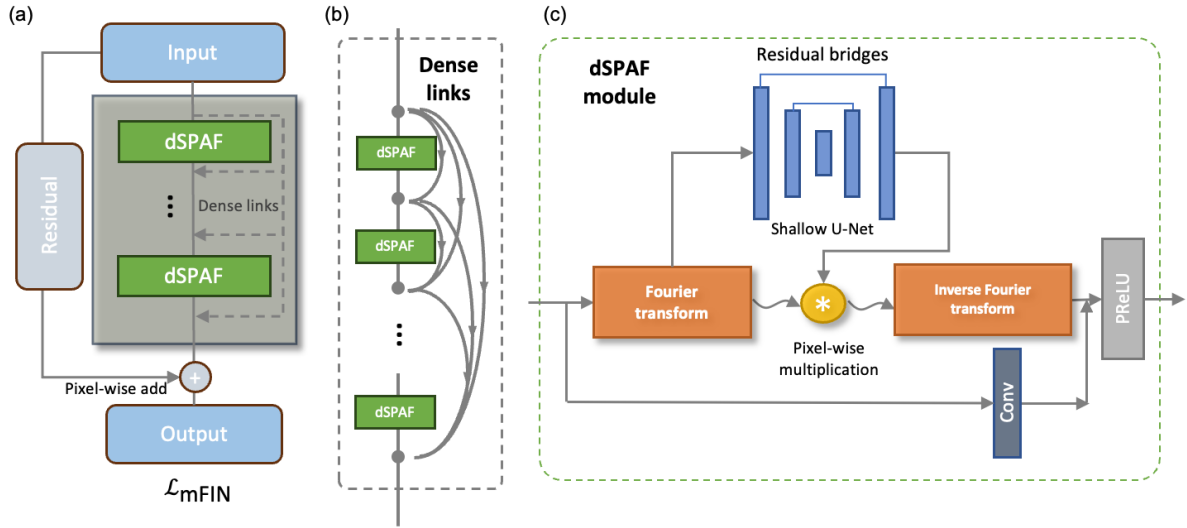

**Figure 7**. Network architecture. (a) Multi-spectral Fourier Imager Network (mFIN). (b) Dense links: each output tensor of the dSPAF group is appended and fed to the subsequent one. (c) Detailed schematic of dSPAF modules. See the Methods section for details.